\documentstyle[12pt,graphicx,amsmath]{article}
\newlength{\pubnumber} 

\catcode`\@=11
\@addtoreset{equation}{section}

\def\section{\@startsection{section}{1}{\z@}{3.5ex plus 1ex minus .2ex}
 {2.3ex plus .2ex}{\large\bf}}
\def\subsection{\@startsection{subsection}{2}{\z@}{2.3ex plus .2ex}
 {2.3ex plus .2ex}{\bf}}

\newcommand{\ba}{\begin{eqnarray}}
\newcommand{\ea}{\end{eqnarray}}
\newcommand{\ob}[1]{\bar{\textbf{#1}}}
\newcommand{\bb}[1]{{\textbf{#1}}}
\newcommand{\hi}[1]{}
\begin{document}

\begin{titlepage}
\samepage{
\setcounter{page}{1}
\rightline{LTH--1030}
\rightline{December 2014}

\vfill
\begin{center}
 {\Large \bf A Light $Z^\prime$ Heterotic--String Derived Model
}
\vspace{1cm}
\vfill {\large
Alon E. Faraggi$^{1}$,
 and
John Rizos$^{2}$}\\
\vspace{1cm}
{\it $^{1}$ Dept.\ of Mathematical Sciences,
             University of Liverpool,
         Liverpool L69 7ZL, UK\\}
\vspace{.05in}
{\it $^{2}$ Department of Physics,
              University of Ioannina, GR45110 Ioannina, Greece\\}
\vspace{.025in}
\end{center}
\vfill
\begin{abstract}

The existence of an extra $Z^\prime$ inspired from heterotic--string theory
at accessible energy scales attracted considerable interest in the 
particle physics literature. Surprisingly, however, the construction 
of heterotic--string derived models that allow for an extra $Z^\prime$ to 
remain unbroken down to low scales has proven to be very difficult. 
The main reason being that the $U(1)$ symmetries that are typically discussed
in the literature are either anomalous or have to be broken at a high
scale to generate light neutrino masses. In this paper we 
use for that purpose
the self duality property under the spinor vector duality, which 
was discovered in free fermionic heterotic string models. 
The chiral massless states in the self--dual models fill 
complete $\mathbf{27}$ representations of $E_6$. The anomaly free gauge
symmetry in the effective low energy field theory
of our string model is $SU(4)_C\times SU(2)_L\times SU(2)_R\times U(1)_\zeta$, 
where $U(1)_\zeta$ is the family universal
$U(1)$ symmetry that descends from $E_6$, 
and is typically anomalous in other free fermionic heterotic--string
models. Our model therefore allows for the existence of 
a low scale $Z^\prime$, which is a combination of $B-L$, $T_{3_L}$ and
$T_{3_R}$. The string model is free of exotic fractionally charged
states in the massless spectrum. It contains exotic $SO(10)$ 
singlet states that carry fractional, non--$E_6$ charge, with respect 
to $U(1)_\zeta$. These non--$E_6$ string states arise in the model
due to the breaking 
of the $E_6$ symmetry by discrete Wilson lines. They 
represent a distinct signature of the string vacua.
%and cannot arise in $E_6$ Grand Unified Theories. 
%%
They may provide viable dark matter candidates. 

\noindent

\end{abstract}
\smallskip}
\end{titlepage}

\setcounter{footnote}{0}

% ========================= DEFINITIONS ===================================
\def\beq{\begin{equation}}
\def\eeq{\end{equation}}
\def\beqn{\begin{eqnarray}}
\def\eeqn{\end{eqnarray}}

\def\no{\noindent }
\def\nolabel{\nonumber }
\def\ie{{\it i.e.}}
\def\eg{{\it e.g.}}
\def\half{{\textstyle{1\over 2}}}
\def\third{{\textstyle {1\over3}}}
\def\quarter{{\textstyle {1\over4}}}
\def\sixth{{\textstyle {1\over6}}}
\def\m{{\tt -}}
\def\p{{\tt +}}

\def\Tr{{\rm Tr}\, }
\def\tr{{\rm tr}\, }

\def\slash#1{#1\hskip-6pt/\hskip6pt}
\def\slk{\slash{k}}
\def\GeV{\,{\rm GeV}}
\def\TeV{\,{\rm TeV}}
\def\y{\,{\rm y}}
\def\SM{Standard--Model }
\def\SUSY{supersymmetry }
\def\SSSM{supersymmetric standard model}
\def\vev#1{\left\langle #1\right\rangle}
\def\l{\langle}
\def\r{\rangle}
\def\o#1{\frac{1}{#1}}

\def\Htw{{\tilde H}}
\def\chibar{{\overline{\chi}}}
\def\qbar{{\overline{q}}}
\def\ibar{{\overline{\imath}}}
\def\jbar{{\overline{\jmath}}}
\def\Hbar{{\overline{H}}}
\def\Qbar{{\overline{Q}}}
\def\abar{{\overline{a}}}
\def\alphabar{{\overline{\alpha}}}
\def\betabar{{\overline{\beta}}}
\def\tautwo{{ \tau_2 }}
\def\thetatwo{{ \vartheta_2 }}
\def\thetathree{{ \vartheta_3 }}
\def\thetafour{{ \vartheta_4 }}
\def\ttwo{{\vartheta_2}}
\def\tthree{{\vartheta_3}}
\def\tfour{{\vartheta_4}}
\def\ti{{\vartheta_i}}
\def\tj{{\vartheta_j}}
\def\tk{{\vartheta_k}}
\def\calF{{\cal F}}
\def\smallmatrix#1#2#3#4{{ {{#1}~{#2}\choose{#3}~{#4}} }}
\def\ab{{\alpha\beta}}
\def\Minv{{ (M^{-1}_\ab)_{ij} }}
\def\bone{{\bf 1}}
\def\ii{{(i)}}
\def\V{{\bf V}}
\def\N{{\bf N}}

% for basis vectors:
\def\b{{\bf b}}
\def\S{{\bf S}}
\def\X{{\bf X}}
\def\I{{\bf I}}
\def\mb{{\mathbf b}}
\def\mS{{\mathbf S}}
\def\mX{{\mathbf X}}
\def\mI{{\mathbf I}}
\def\balpha{{\mathbf \alpha}}
\def\bbeta{{\mathbf \beta}}
\def\bgamma{{\mathbf \gamma}}
\def\bxi{{\mathbf \xi}}

\def\t#1#2{{ \Theta\left\lbrack \matrix{ {#1}\cr {#2}\cr }\right\rbrack }}
\def\C#1#2{{ C\left\lbrack \matrix{ {#1}\cr {#2}\cr }\right\rbrack }}
\def\tp#1#2{{ \Theta'\left\lbrack \matrix{ {#1}\cr {#2}\cr }\right\rbrack }}
\def\tpp#1#2{{ \Theta''\left\lbrack \matrix{ {#1}\cr {#2}\cr }\right\rbrack }}
\def\l{\langle}
\def\r{\rangle}
\newcommand{\cc}[2]{c{#1\atopwithdelims[]#2}}
\newcommand{\nn}{\nonumber}

%================== BLACKBOARD BOLD CHARACTERS ==============================

\def\inbar{\,\vrule height1.5ex width.4pt depth0pt}

\def\IC{\relax\hbox{$\inbar\kern-.3em{\rm C}$}}
\def\IQ{\relax\hbox{$\inbar\kern-.3em{\rm Q}$}}
\def\IR{\relax{\rm I\kern-.18em R}}
 \font\cmss=cmss10 \font\cmsss=cmss10 at 7pt
\def\IZ{\relax\ifmmode\mathchoice
 {\hbox{\cmss Z\kern-.4em Z}}{\hbox{\cmss Z\kern-.4em Z}}
 {\lower.9pt\hbox{\cmsss Z\kern-.4em Z}}
 {\lower1.2pt\hbox{\cmsss Z\kern-.4em Z}}\else{\cmss Z\kern-.4em Z}\fi}

%========================================================================
%          MACROS FOR REFERENCES
%========================================================================
\def\AEF{A.E. Faraggi}
\def\JHEP#1#2#3{{\it JHEP}\/ {\bf #1} (#2) #3}
\def\NPB#1#2#3{{\it Nucl.\ Phys.}\/ {\bf B#1} (#2) #3}
\def\PLB#1#2#3{{\it Phys.\ Lett.}\/ {\bf B#1} (#2) #3}
\def\PRD#1#2#3{{\it Phys.\ Rev.}\/ {\bf D#1} (#2) #3}
\def\PRL#1#2#3{{\it Phys.\ Rev.\ Lett.}\/ {\bf #1} (#2) #3}
\def\PRT#1#2#3{{\it Phys.\ Rep.}\/ {\bf#1} (#2) #3}
\def\MODA#1#2#3{{\it Mod.\ Phys.\ Lett.}\/ {\bf A#1} (#2) #3}
\def\RMP#1#2#3{{\it Rev.\ Mod.\ Phys.}\/ {\bf #1} (#2) #3}
\def\IJMP#1#2#3{{\it Int.\ J.\ Mod.\ Phys.}\/ {\bf A#1} (#2) #3}
\def\nuvc#1#2#3{{\it Nuovo Cimento}\/ {\bf #1A} (#2) #3}
\def\RPP#1#2#3{{\it Rept.\ Prog.\ Phys.}\/ {\bf #1} (#2) #3}
\def\EJP#1#2#3{{\it Eur.\ Phys.\ Jour.}\/ {\bf C#1} (#2) #3}
\def\etal{{\it et al\/}}

%==========================================================================
\hyphenation{su-per-sym-met-ric non-su-per-sym-met-ric}
\hyphenation{space-time-super-sym-met-ric}
\hyphenation{mod-u-lar mod-u-lar--in-var-i-ant}
%==========================================================================

%============================== SECTION 1 ============================

\setcounter{footnote}{0}
\section{Introduction}

The consistency conditions of string theory necessitate the 
existence of additional gauge degrees of freedom beyond those 
that are observed in the Standard Model. 
Experimental observation of extra gauge degrees 
of freedom in contemporary experiment will lend 
evidence for the extra gauge degrees of freedom
predicted in string theory. 
The Standard Model states may be neutral under some of these 
degrees of freedom and charged with respect to some others. 
The neutral sector is dubbed the hidden sector, 
and typically consists of a rank eight gauge group.
The observable sector of the heterotic--string correspond 
to a rank eight group, whereas the Standard Model 
utilises four of these degrees of freedom. 
Naturally, the experimental
signatures of extra vector bosons arising in the 
hidden and observable sectors will be markedly 
different. In this paper we consider the case
with an extra vector boson arising in the observable 
sector.  

Extra $U(1)$ gauge symmetries in string theory have
been of interest since the mid--eighties and occupy a 
significant number of papers that use effective
field theory methods to study their phenomenological 
implications 
\cite{zphistory,zpbminusl, zpfff,fm1}. 
Surprisingly, however, the construction
of viable heterotic--string models that admit an 
additional observable $U(1)$ vector boson that may 
remain unbroken down to low energies, has proven 
to be very difficult, for a variety of phenomenological
restrictions. In fact, to date there does not exist a free fermionic
heterotic--string derived model that allows an extra observable 
$U(1)$ symmetry to remain unbroken down to low scales.

One issue that must be addressed is that of simultaneously
suppressing proton decay mediating operators, while allowing 
for a mechanism that suppresses the left--handed neutrino masses
\cite{zpfff}. 
Embedding the Standard Model in $SO(10)$ extends the rank of the
Standard Model gauge group by one. Hence giving rise to 
an extra $U(1)$, which is a combination of $B-L$, baryon minus 
lepton number, and $T_{3_R}$, the diagonal generator of $SU(2)_R$. 
The existence of this extra $U(1)$ at low scales was already entertained
in the late eighties \cite{zpbminusl}. 
%, and has been a subject of some recent renewed interest 
%\cite{ovrut}. 
The caveat is that since the lepton number 
is gauged the extra $U(1)$ symmetry forbids the formation 
of Majorana mass terms for the right--handed neutrinos.
On the other hand, the underlying $SO(10)$ symmetry 
dictates the equality of the top--quark and tau--neutrino
Yukawa couplings, and hence the equality of the tau--neutrino
Dirac mass term and the top quark mass. Preserving $U(1)_{B-L}$ unbroken 
down to the TeV scale, entails a low seesaw scale and 
a tau neutrino mass scale of the order of $O(10{\rm MeV})$
\cite{tauneutrinomass}. 
Ensuring neutrino masses below the eV scale necessitates that
$U(1)_{B-L}$ is broken at a scale of order
$O(10^{15}{\rm GeV})$ \cite{zpfff}. 

Another problem arises from the fact that in many string
models the additional family universal $U(1)$ symmetries, 
which are traditionally studied in string inspired
constructions, are anomalous and are not viable at low scales.
The reason is the particular symmetry breaking pattern that
is realised in many of the quasi--realistic free fermionic
heterotic--string models 
\cite{fsu5,slm,alr}. It can be seen to arise from the
breaking of $E_6\rightarrow SO(10)\times U(1)_\zeta$, which results 
in $U(1)_\zeta$ being anomalous, since the chiral matter 
resides in incomplete $E_6$ representations \cite{cleaverau1}. 
The left--right symmetric 
heterotic string models \cite{lrs}
circumvent this symmetry breaking
pattern and do produce anomaly free models. 
On the other hand, string inspired constructions that utilise 
the $U(1)_\zeta$ charge assignments of the left--right symmetric 
heterotic string models \cite{fm1}
%extra anomaly free $U(1)$s 
disagree with the gauge 
coupling data \cite{mehtathree}. The reason is that the charges of the 
Standard Model states under the extra $U(1)$ do not 
admit an $E_6$ embedding, which is a necessary ingredient
for accommodating the gauge coupling data \cite{mehtathree}. 

The challenge is therefore to construct three generation 
string models that allow for an extra family universal 
$U(1)$ symmetry with $E_6$ embedding of its chiral charges.
The $E_6$ symmetry is broken directly at the string level 
and is not manifested at low scales. The fact that the chiral
spectrum must be anomaly free entails that the chiral generations
must come in complete $E_6$ multiplets. 
In this paper we use for that purpose 
the spinor vector duality that was observed 
in $Z_2\times Z_2$ heterotic--string orbifolds with 
$SO(10)$ GUT symmetry \cite{fkr,spinvecdual, aft, ffmt}.
The spinor--vector duality entails that for every 
string vacuum with a number of $\mathbf{16}\oplus\overline{{16}}$, and 
a number $\mathbf{10}$ representations of $SO(10)$, there exist another
vacuum in which the two numbers are interchanged. 
The spinor vector duality was first noted in 
the classification of free fermion $SO(10)$ models
\cite{fkr,spinvecdual} in terms of the Generalised GSO 
projection coefficients. It was subsequently discussed 
in terms of discrete torsion in orbifold models \cite{aft, ffmt}.
It was shown to arise generally from the breaking of the world--sheet
supersymmetry from $(2,2) \rightarrow (2,0)$, and is induced
by the spectral flow operator of the right--moving 
world--sheet supersymmetry \cite{ffmt}. 
A special class of models are the self--dual models under
the spinor--vector duality, which contain an 
equal number of $\mathbf{16}\oplus\overline{{16}}$ and $\mathbf10$, 
representations of $SO(10)$. In the self--dual models 
$U(1)_\zeta$ may be anomaly free without enhancement of 
the gauge symmetry to $E_6$. The reason being 
that the spinorial and vectorial states that form complete 
$E_6$ representations are obtained from different
fixed points of the underlying $Z_2\times Z_2$ orbifold. 
The next step in our construction is to add a basis
vector that breaks the $SO(10)$ symmetry to the Pati--Salam 
subgroup \cite{ps, alr}, while maintaining the spinor--vector
self--duality. We present an exemplary three generation model 
with these characteristics, which is free of exotic fractionally
charged states, and contains the Higgs states necessary for 
realistic phenomenology. An interesting property of the model
is that while it is free of $SO(10)$ exotic states, it 
contains states that carry exotic charges with respect to 
$U(1)_\zeta$, {\it i.e.} states that are not descending
from $E_6$ representations. Such states are therefore 
signature of the string models.
% and do not appear in the 
%corresponding GUT constructions. 
%%
Furthermore, these states fall into the general
category of Wilsonian matter states, considered 
in ref. \cite{wilsonian}, and therefore may provide
viable dark matter candidates. The 
reason being that breaking the $U(1)_{Z^\prime}$ with $E_6$ states
leaves a remnant discrete symmetry that forbids the 
decay of the exotic states to the Standard Model states,
which carry standard $E_6$ charges. 

\section{A String derived extra $U(1)$ model}\label{model}

Our challenge is to construct three generation 
heterotic--string models that allow for an extra family universal 
$U(1)$ symmetry with $E_6$ embedding of the chiral charges.
The $E_6$ symmetry is broken directly at the string level. 
The fact that the chiral spectrum must be anomaly free 
entails that the chiral generations
must come in complete $E_6$ multiplets. 
In refs \cite{mehtathree}
and \cite{mehtafour} the construction of
$SO(6)\times SO(4)\times U(1)_\zeta$ and 
$SU(3)_C\times SU(2)_W\times U(1)_{B-L}\times U(1)_{T_{3_R}}\times U(1)_\zeta$
models was outlined. In both cases the symmetry is broken spontaneously
to $SU(3)_C\times SU(2)_W\times U(1)_Y\times U(1)_{Z^\prime}$
by the vacuum expectation value of the Standard Model
singlet in the spinorial $\mathbf{16}$ representation of $SO(10)$.
The outline of these constructions goes as follows. 
The spacetime vector bosons that produce
the observable $E_8$ gauge symmetry in free fermion models 
are obtained from two sectors. The first is the untwisted sector 
and the second is the $x$--sector \cite{enahe}. In the decomposition of
$E_8\rightarrow SO(16)$, the adjoint representation 
decomposes as $\mathbf{248}\rightarrow \mathbf{120}+\mathbf{128}$, where the 
$\mathbf{120}$ representation is obtained from the untwisted 
sector, whereas the $\mathbf{128}$ is obtained from the $x$--sector. 
In many of the existing quasi--realistic free fermionic 
models the states from the $x$--sector are projected 
out. The consequence is that $U(1)_\zeta$ and consequently
$U(1)_{Z^\prime}$ is anomalous. The key to the 
proposals in \cite{mehtathree, mehtafour} is
to construct models in which some of the vector bosons
from the $x$--sector are retained in the spectrum and 
enhance the untwisted gauge symmetry. 

An explicit realisation of such a construction
is the $SU(6)\times SU(2)$ model of \cite{lauraivan}.
The caveat with this model is that the only 
scalar states available to
break the gauge symmetry down to the Standard Model
are obtained from the $\mathbf{27}$ of $E_6$. It is therefore 
impossible to break the symmetry down to the 
Standard Model, while maintaining an unbroken 
extra $U(1)$ symmetry. The reason being 
that this model requires two stages of non--Abelian
symmetry breaking. The first being the breaking of 
$SU(6)\times SU(2)$ to
either the Pati--Salam or flipped $SU(5)$ subgroups,
and the second being the breaking of these subgroups to 
the Standard Model. The strategy proposed in ref. 
\cite{mehtathree, mehtafour} is therefore to 
construct similar models, but in which the enhancement
of the untwisted gauge group is to 
$SO(6)\times SO(4)\times U(1)_\zeta$ and 
$SU(3)_C\times SU(2)_W\times U(1)_{B-L}\times U(1)_{T_{3_R}}\times U(1)_\zeta$, 
respectively, rather than to $SU(6)\times SU(2)$. 
However, explicit string derived models that realise 
this construction were not presented in refs. \cite{mehtathree, mehtafour}. 

In this paper we adopt an alternative construction that exploits the
spinor--vector duality observed in free fermionic models in ref.
\cite{fkr,spinvecdual}. The spinor--vector duality exchanges spinorial
$\mathbf{16}$ representations of $SO(10)$ with vectorial 
$\mathbf{10}$ representations
in the twisted sectors. For every vacuum with a total number
of $\mathbf{16}\oplus{\overline{{16}}}$ multiplets and a number of
$\mathbf{10}$ multiplets, there exist a dual vacuum in which the 
two numbers are interchanged. The spinor--vector 
duality can be proved analytically in terms of the 
free fermion Generalised GSO (GGSO) phases of
the one--loop partition function \cite{spinvecdual}, 
or in terms of discrete torsions in an orbifold representation 
\cite{aft,ffmt}. 
It can be seen to arise due to the breaking of the 
$N=2\rightarrow N=0$ world--sheet supersymmetry
in the right--moving bosonic side of the heterotic--string.
With $N=2$ world--sheet supersymmetry the
$SO(10)\times U(1)$ GUT symmetry is enhanced
to $E_6$. The chiral multiplets reside in
the $\mathbf{27}$ and $\overline{\mathbf{27}}$ representations 
of $E_6$, which decompose as 
$\mathbf{27}=\mathbf{16}_{+{1\over2}}+\mathbf{10}_{-1}+\mathbf{1}_{+2}$
and $\overline{\mathbf{27}}=\overline{\mathbf{16}}_{-{1\over2}}+\mathbf{10}_{+1}+\mathbf{1}_{-2}$,
respectively, under $SO(10)\times U(1)_\zeta$. 
When the symmetry is enhanced to $E_6$ 
the total number of $\mathbf{16}\oplus{\overline{\mathbf{16}}}$ representations
is equal to the total number of vectorial 
$\mathbf{10}$ representations. 
Hence, this case is self--dual under the spinor--vector duality. 
In this case the spectral flow generator on the bosonic
side exchanges between the multiplets that are embedded 
in the $E_6$ representations. Breaking the $N=2$ world--sheet
supersymmetry to $N=0$ induces the $E_6\rightarrow SO(10)\times U(1)_\zeta$
breaking. In this case the spectral flow operator induces the 
spinor--vector duality map between the dual vacua \cite{ffmt}. 
Since, the $E_6$ symmetry is broken, the chiral spectrum 
resides in incomplete $E_6$ multiplets, and $U(1)_\zeta$ is,
in general, anomalous. 
A special class of models are the $N=0$ self--dual models under the 
spinor--vector duality map. In these models the $E_6$ symmetry is 
broken to $SO(10)\times U(1)$. However the total number
of spinor plus anti--spinor representations is equal 
to the total number of vectorial representations. 
Hence, these models produce complete $E_6$ multiplets,
but the gauge symmetry is not enhanced to $E_6$. 
This is possible if the different components of the 
$E_6$ multiplets are obtained from different fixed points
of the underlying $Z_2\times Z_2$ orbifold. 
Obtaining the spinorial and vectorial components 
at the same fixed point would necessarily imply that the
$SO(10)\times U(1)$ symmetry is enhanced to $E_6$. However, 
if the spinorial and vectorial components are obtained
at different fixed points the symmetry is not enhanced. 
The chiral spectrum in the self--dual models may therefore
arise in complete $E_6$ multiplets, with anomaly free 
$U(1)_\zeta$, but without enhanced $E_6$ symmetry. The next stage
in our construction is to break the $SO(10)\times U(1)_\zeta$ 
symmetry to $SO(6)\times SO(4)\times U(1)_\zeta$, while maintaining 
the spinor--vector self--duality.

We use the free fermionic formulation of the heterotic string in 
four dimensions \cite{fff}
to construct our string derived model. In this formulation
all the degrees of freedom required to cancel the world--sheet 
conformal anomaly are represented in terms of free fermions propagating
on the string world--sheet. These fermions pick up a phase under
parallel transport around the non--contractible
loops of the worldsheet torus. The free fermion
heterotic string models are fully described in terms 
of the boundary condition basis vectors
$v_i,i=1,\dots,N$
$$v_i=\left\{v_i(f_1),v_i(f_{2}),v_i(f_{3}))\dots\right\},$$
for the 64 world--sheet real fermions $f_j$ \cite{fff}, 
and the associated 
one--loop GGSO coefficients
$ \cc{v_i}{v_j}$. Taking all possible combinations of 
the basis vectors
\begin{equation}
\eta = \sum N_i v_i,\ \  N_i =0,1
\end{equation}
generates a finite additive group $\Xi$. 
The physical states in each sector $\eta\in\Xi$ are 
obtained by acting on the vacuum with fermionic and 
bosonic oscillators and by imposing the
GGSO projections
\begin{equation}\label{eq:gso}
e^{i\pi v_i\cdot F_S} |S> = \delta_{S}\ \cc{S}{v_i} |S>,
\end{equation}
where $\delta_{S}=\pm1$ is the spacetime spin 
statistics index and
$F_S$ is a fermion number operator.
In the usual notation
the sixty--four worldsheet fermions in the 
light--cone gauge are: 
$\psi^\mu, \chi^i,y^i, \omega^i, i=1,\dots,6$ (left-movers) and
$\bar{y}^i,\bar{\omega}^i, i=1,\dots,6$,
$\psi^A, A=1,\dots,5$, $\bar{\eta}^B, B=1,2,3$, $\bar{\phi}^\alpha,
\alpha=1,\ldots,8$ (right-movers).
Further details of 
the formalism and notation that we use in this 
paper are found in the literature 
\cite{fff,fsu5,slm,alr,enahe,fknr,fkr,psclass}.

Our string model is constructed by using the methods
developed in \cite{gkr} for the classification of
type IIB superstrings, and in 
\cite{fknr} for the classification 
heterotic--string vacua with an unbroken $SO(10)$ symmetry.
It was adapted in \cite{psclass} for the classification of 
Pati--Salam vacua, and in \cite{fsu5class} for the 
classification of flipped $SU(5)$ vacua. In this classification
method the set of basis vectors is fixed and the enumeration of the
models is achieved by varying GGSO phases.
The set of basis vectors that we use here is
identical to the one used in the 
classification of Pati--Salam vacua in ref \cite{psclass} 
and is given by a set of thirteen basis vectors
$
B=\{v_1,v_2,\dots,v_{13}\},
$
where
\begin{eqnarray}
v_1=1&=&\{\psi^\mu,\
\chi^{1,\dots,6},y^{1,\dots,6}, \omega^{1,\dots,6}| \nonumber\\
& & ~~~\bar{y}^{1,\dots,6},\bar{\omega}^{1,\dots,6},
\bar{\eta}^{1,2,3},
\bar{\psi}^{1,\dots,5},\bar{\phi}^{1,\dots,8}\},\nonumber\\
v_2=S&=&\{\psi^\mu,\chi^{1,\dots,6}\},\nonumber\\
v_{2+i}=e_i&=&\{y^{i},\omega^{i}|\bar{y}^i,\bar{\omega}^i\}, \
i=1,\dots,6,\nonumber\\
v_{9}=b_1&=&\{\chi^{34},\chi^{56},y^{34},y^{56}|\bar{y}^{34},
\bar{y}^{56},\bar{\eta}^1,\bar{\psi}^{1,\dots,5}\},\label{basis}\\
v_{10}=b_2&=&\{\chi^{12},\chi^{56},y^{12},y^{56}|\bar{y}^{12},
\bar{y}^{56},\bar{\eta}^2,\bar{\psi}^{1,\dots,5}\},\nonumber\\
v_{11}=z_1&=&\{\bar{\phi}^{1,\dots,4}\},\nonumber\\
v_{12}=z_2&=&\{\bar{\phi}^{5,\dots,8}\},\nonumber\\
v_{13}=\alpha &=& \{\bar{\psi}^{4,5},\bar{\phi}^{1,2}\}.\nonumber
\end{eqnarray}
In the notation used in eq. (\ref{basis})
the fermions appearing in the curly brackets are periodic,
whereas those that do not appear are antiperiodic. 
The untwisted gauge symmetry generated by this set 
is 
\beqn
{\rm observable} ~: &~~~~~~~~SO(6)\times SO(4) \times U(1)^3 \nonumber\\
{\rm hidden}     ~: &~~SO(4)^2\times SO(8)~~~~             \nonumber
\eeqn
Additional spacetime vector bosons may arise from 
the sectors 
%%%%%%%%%%%%%%%%%%%%%%%%%%%%%%%%%%%%%%%
\begin{equation}
\mathbf{G} =
\left\{ \begin{array}{cccccc}
z_1          ,&
z_2          ,&
\alpha       ,&
\alpha + z_1 ,&
              &
                \cr
x            ,&
z_1 + z_2    ,&
\alpha + z_2 ,&
\alpha + z_1 + z_2,&
\alpha + x ,&
\alpha + x + z_1
\end{array} \right\} \label{stvsectors}
\end{equation}
%%%%%%%%%%%%%%%%%%%%%%%%%%%%%%%%%%%%%%
and enhance the four dimensional gauge group.
In Eq. (\ref{stvsectors})
we defined the combination $$x=1+S+\sum_{i=1}^6 e_i+z_1+z_2,$$
that may enhance the observable $SO(16)$ gauge group to $E_8$.
For suitable choices of the GGSO phases the spacetime 
gauge bosons arising in the sectors of eq. (\ref{stvsectors})
are projected out, and the gauge symmetry is generated solely
by the vector bosons arising in the untwisted sector. 
A suitable choice of the GGSO projection coefficients 
guarantees the existence of $N=1$ spacetime supersymmetry.

The matter states in the Pati--Salam heterotic--string
models are embedded in 
$SU(4)\times{SU(2)}_L\times{SU(2)}_R$
representations as follows:
\begin{align}
 {F}_L\left({\bf4},{\bf2},{\bf1}\right)     &\rightarrow
     q\left({\bf3},{\bf2},-\frac 16\right) + 
     \ell{\left({\bf1},{\bf2},\frac 12\right)}
     \nonumber\\
\bar{F}_R\left({\bf\bar 4},{\bf1},{\bf2}\right)&\rightarrow   
  u^c\left({\bf\bar 3},{\bf1},\frac 23\right)+
  d^c\left({\bf\bar 3},{\bf1},-\frac 13\right)+
  e^c\left({\bf1},{\bf1},-1)+
  \nu^c({\bf1},{\bf1},0\right)
     \nonumber\\
h({\bf1},{\bf2},{\bf2}) &\rightarrow   
  h^d\left({\bf1},{\bf2},\frac 12\right) + 
  h^u\left({\bf1},{\bf2},-\frac 12\right)
    \nonumber\\
D\left({\bf6},{\bf1},1\right)  &\rightarrow   
d_3\left({\bf3},{\bf1},\frac 13\right) +
\bar{d}_3\left({\bf\bar 3},{\bf1},-\frac 13\right). 
                             \nonumber
\end{align}
Here ${\bar F}_R$ and $F_L$ contain one Standard Model generation;
$h^u$ and $h^d$ are electroweak Higgs doublets; 
and $D$ contains vector--like colour triplets.
The Pati--Salam breaking Higgs fields, decomposed 
in terms of the Standard Model group factors, are given by:
\begin{align}
	\bar{H}({\bf\bar 4},{\bf1},{\bf2})   & \rightarrow   u^c_H\left({\bf\bar 3},
{\bf1},\frac 23\right)+d^c_H\left({\bf\bar 3},{\bf1},-\frac 13\right)+
                            \nu^c_H\left({\bf1},{\bf1},0\right)+
                             e^c_H\left({\bf1},{\bf1},-1\right)
                             \nonumber \\
	{H}\left({\bf4},{\bf1},{\bf2}\right) & \rightarrow  u_H\left({\bf3},{\bf1},-\frac
23\right)+d_H\left({\bf3},{\bf1},\frac 13\right)+
              \nu_H\left({\bf1},{\bf1},0\right)+ e_H\left({\bf1},{\bf1},1\right)\nn
\end{align}
The electric charge in the Pati--Salam models is given by:
\beq
Q_{em} = {1\over\sqrt{6}}T_{15}+{1\over2}T_{3_L}+{1\over2}T_{3_R}
\eeq
where $T_{15}$ is the diagonal generator of $SU(4)$ and
$T_{3_L}$, $T_{3_R}$
are the diagonal generators of $SU(2)_L$, $SU(2)_R$, respectively.

The next ingredient required to define the string model are 
the GGSO projection coefficients that are obtained
from the one--loop partition function
$\cc{v_i}{v_j}$, spanning a $13\times 13$ matrix.
Modular invariance constraints imply that only the 
elements with $i>j$ are independent. 
There are therefore a priori 78 independent coefficients 
of which 11 are fixed by the requirement that
the models possess $N=1$ spacetime supersymmetry.
The number of independent phases is reduced further 
to 51, by the requirement that only untwisted 
spacetime vector bosons are retained in the massless
spectrum. Each distinct configuration of the GGSO phases
corresponds to a distinct model, where some degeneracy in 
some of the phenomenological properties may still exist. 
A statistical analysis over the entire space of models 
was presented in \cite{psclass}. Here our interest
is in the particular class of models that preserve the 
spinor--vector self--duality and that admit the 
additional anomaly free $U(1)_\zeta$ symmetry in the 
observable sector. The phases displayed in eq. (\ref{BigMatrix}),

\beq \label{BigMatrix}  (v_i|v_j)\ \ =\ \ \bordermatrix{
      & 1& S&e_1&e_2&e_3&e_4&e_5&e_6&b_1&b_2&z_1&z_2&\alpha\cr
 1    & 1& 1&  1&  1&  1&  1&  1&  1&  1&  1&  1&  1&      1\cr
S     & 1& 1&  1&  1&  1&  1&  1&  1&  1&  1&  1&  1&      1\cr
e_1   & 1& 1&  0&  0&  0&  0&  0&  0&  0&  0&  0&  0&      1\cr
e_2   & 1& 1&  0&  0&  0&  0&  0&  1&  0&  0&  0&  1&      0\cr
e_3   & 1& 1&  0&  0&  0&  1&  0&  0&  0&  0&  0&  1&      1\cr
e_4   & 1& 1&  0&  0&  1&  0&  0&  0&  0&  0&  1&  0&      0\cr
e_5   & 1& 1&  0&  0&  0&  0&  0&  1&  0&  0&  0&  1&      1\cr
e_6   & 1& 1&  0&  1&  0&  0&  1&  0&  0&  0&  1&  0&      0\cr
b_1   & 1& 0&  0&  0&  0&  0&  0&  0&  1&  1&  0&  0&      0\cr
b_2   & 1& 0&  0&  0&  0&  0&  0&  0&  1&  1&  0&  0&      1\cr
z_1   & 1& 1&  0&  0&  0&  1&  0&  1&  0&  0&  1&  1&      0\cr
z_2   & 1& 1&  0&  1&  1&  0&  1&  0&  0&  0&  1&  1&      0\cr
\alpha& 1& 1&  1&  0&  1&  0&  1&  0&  1&  0&  1&  0&      1\cr
  }
\eeq
where we introduced the notation
$\cc{v_i}{v_j} = e^{i\pi (v_i|v_j)}$,
represent a specific example of a heterotic--string model in 
this class. We estimate the existence of some $2\times 10^5$ 
models with similar properties. The model was obtained 
using a fishing algorithm to extract a specific 
configuration with particular phenomenological properties
\cite{psclass}. An alternative method is to use the genetic algorithm
proposed in ref. \cite{genetic}.

%new from John
\begin{table}[!h]
\noindent
{\small
\openup\jot
\begin{tabular}{|l|l|c|c|c|c||c|}
\hline
sector&field&$SU(4)\times{SU(2)}_L\times{SU(2)}_R$&${U(1)}_1$&${U(1)}_2$&${U(1)}_3$&$U(1)_\zeta$\\
%\hline
\hline
$S$&$D_1$&$(\bb{6},\bb{1},\bb{1})$&$-1$&$\hphantom{+}0$&$\hphantom{+}0$&$-1$\\
&$D_2$&$(\bb{6},\bb{1},\bb{1})$&$\hphantom{+}0$&$-1$&$\hphantom{+}0$&$-1$\\
&$D_3$&$(\bb{6},\bb{1},\bb{1})$&$\hphantom{+}0$&$\hphantom{+}0$&$-1$&$-1$\\
&$\bar{D}_1$&$(\bb{6},\bb{1},\bb{1})$&$\hphantom{+}1$&$\hphantom{+}0$&$\hphantom{+}0$&{+}1\\
&$\bar{D}_2$&$(\bb{6},\bb{1},\bb{1})$&$\hphantom{+}0$&$\hphantom{+}1$&$\hphantom{+}0$&{+}1\\
&$\bar{D}_3$&$(\bb{6},\bb{1},\bb{1})$&$\hphantom{+}0$&$\hphantom{+}0$&$\hphantom{+}1$&{+}1\\
&$\Phi_{12}$&$(\bb{1},\bb{1},\bb{1})$&$+1$&$+1$&$\hphantom{+}0$&${+}2$\\
&$\bar{\Phi}_{12}$&$(\bb{1},\bb{1},\bb{1})$&$-1$&$-1$&$\hphantom{+}0$&$-2$\\
&$\Phi_{13}$&$(\bb{1},\bb{1},\bb{1})$&$+1$&$\hphantom{+}0$&$+1$&${+}2$\\
&$\bar{\Phi}_{13}$&$(\bb{1},\bb{1},\bb{1})$&$-1$&$\hphantom{+}0$&$-1$&$-2$\\
&$\Phi_{23}$&$(\bb{1},\bb{1},\bb{1})$&$\hphantom{+}0$&$+1$&$+1$&${+}2$\\
&$\bar{\Phi}_{23}$&$(\bb{1},\bb{1},\bb{1})$&$\hphantom{+}0$&$-1$&$-1$&$-2$\\
&$\Phi_{12}^{-}$&$(\bb{1},\bb{1},\bb{1})$&$+1$&$-1$&$\hphantom{+}0$&$\hphantom{+}0$\\
&$\bar{\Phi}_{12}^{-}$&$(\bb{1},\bb{1},\bb{1})$&$-1$&$+1$&$\hphantom{+}0$&$\hphantom{+}0$\\
&$\Phi_{13}^-$&$(\bb{1},\bb{1},\bb{1})$&$+1$&$\hphantom{+}0$&$-1$&$\hphantom{+}0$\\
&$\bar{\Phi}_{13}^-$&$(\bb{1},\bb{1},\bb{1})$&$-1$&$\hphantom{+}0$&$+1$&$\hphantom{+}0$\\
&$\Phi_i,i=1,\dots,6$&$(\bb{1},\bb{1},\bb{1})$&$\hphantom{+}0$&$\hphantom{+}0$&$\hphantom{+}0$&$\hphantom{+}0$\\
&$\Phi_{23}^-$&$(\bb{1},\bb{1},\bb{1})$&$\hphantom{+}0$&$+1$&$-1$&$\hphantom{+}0$\\
&$\bar{\Phi}_{23}^-$&$(\bb{1},\bb{1},\bb{1})$&$\hphantom{+}0$&$-1$&$+1$&$\hphantom{+}0$\\
\hline
\end{tabular}
}
\caption{\label{tablea}\it
The untwisted matter states and
$SU(4)\times{SU(2)}_L\times{SU(2)}_R\times{U(1)}^3$ charges. }
\end{table}

In tables \ref{tablea}, \ref{tableb} and \ref{tablec} we display the 
entire massless spectrum that arise in the model generated 
by the set of GGSO phases in eq. (\ref{BigMatrix}).
The vector combination defined in the tables as $b_3=b_1+b_2+x$ 
correspond to the third twisted plane of the $Z_2\times Z_2$ 
orbifold.
 
In addition to the spacetime vector bosons that generate the 
four dimensional gauge group
the untwisted sector gives rise to  
three pairs of $SU(4)$ sextets; six pairs of
$SO(10)\times E_8$ singlets that are charged with respect to 
$U(1)$ symmetries; and six states that are neutral under the 
entire four dimensional gauge group. These states are displayed
in table \ref{tablea}. The states arising from the untwisted sector
are identical to all the Pati--Salam free fermionic models that use
the basis set given in eq. (\ref{basis}) since the projections
of the untwisted set only depend on the basis vectors and are
independent of the choice of GGSO projection coefficients given
in eq. (\ref{BigMatrix}).

The twisted sectors matter states obtained in the string 
model of eq. (\ref{BigMatrix}) generate the needed 
states for viable phenomenology. The massless spectrum contains 
three chiral generations; 
one pair of heavy Higgs states to break the Pati--Salam gauge symmetry;
three light Higgs bi--doublets that can be used to break the electroweak
symmetry and generate viable fermion mass spectrum; the twisted spectrum
contains five sextet states of $SO(6)$, where at least one is required 
for the missing partner mechanism. The massless spectrum is
completely free of exotic fractionally charged states that are 
endemic in heterotic string vacua \cite{ww,bert,fc, wilsonian}. 
Additionally, the spectrum contains a number of $SO(10)\times U(1)_\zeta$ 
singlet states, that can be used to produce a supersymmetric vacuum 
along $F$-- and $D$--flat directions. 
Some of these states transform in non--trivial representations
of the hidden sector gauge group. From table \ref{tableb} 
it is seen that the spinor--vector self--duality at the 
$SO(10)$ level is preserved in the Pati--Salam model. 

Several observations can be noted from the 
twisted sectors states displayed in tables \ref{tableb} and \ref{tablec}.
First, it is seen that the chiral representations indeed form 
complete $\mathbf{27}$ representations and consequently $U(1)_\zeta$ is 
anomaly free. The string model contains two anomalous $U(1)$ with
\beq
{\rm Tr}U(1)_1= 36 ~~~~~~~{\rm and}~~~~~~~{\rm Tr}U(1)_3= -36,
\label{u1u3}
\eeq
where the $U(1)_{1,2,3}$ symmetries are generated by 
the right--moving complex worldsheet fermions ${\bar\eta}^{1,2,3}$.  
Hence, the two combinations 
\beqn
U(1)_\zeta & = & U(1)_1 + U(1)_2 + U(1)_3 \label{u1zeta}\\
U(1)_{2^\prime} & = & U(1)_1 -2 U(1)_2 + U(1)_3 \label{u2prime}
\eeqn
are anomaly free, whereas the combination 
\beq
U(1)_A  = U(1)_1 - U(1)_3 \label{u2a}
\eeq
is anomalous. The anomalous $U(1)$ symmetry generates 
a Fayet--Iliopoulos term that breaks supersymmetry
near the Planck scale \cite{dsw}. Supersymmetry can be restored along 
a flat directions by assigning a VEV to some $SO(10)\times U(1)_\zeta$ 
singlet fields in the string massless spectrum, for example by giving 
a VEV to ${\bar\Phi}^-_{13}$. Assigning a VEV to the heavy Higgs field 
that breaks the Pati--Salam symmetry leaves unbroken the 
weak hypercharge combination 
\beq
U(1)_Y={1\over3}U_C+{1\over2}U_L, 
\nonumber
\eeq
and the $Z^\prime$ combination given by
\beq
U(1)_{Z^\prime}={1\over5}U_C-{1\over5}U_L-U_\zeta,
\label{uzprime}
\eeq
where we used the $U(1)$ definitions traditionally used 
in free fermionic models $U_c=3/2U_{B-L}$ and $U_L=2 T_{3_R}$. 

As noted from table \ref{tableb} the $\chi^+_i$ states with $i=1,\cdots,5$
correspond to the $SO(10)$ singlet in the $27$ representation of $E_6$. 
The corresponding $\chi^-_i$ states correspond to the twisted moduli
\cite{enahe,modulifixing}. 
Thus, contrary to the cases \cite{modulifixing} in which the
twisted moduli are projected out, in the self--dual model 
of eq. (\ref{BigMatrix}) they are retained.  
The states $\zeta_a$ and ${\bar\zeta}_a$ with $a=1, \cdots, 11$ 
are neutral under $SO(10)\times U(1)_\zeta$ and can therefore 
get non--trivial VEVs along supersymmetric $F$-- and $D$--flat directions. 

A particularly interesting class of states are the states 
$\phi_{1,2}$ and ${\bar\phi}_{1,2}$. These states are $SO(10)$ singlets
and are charged with respect to $U(1)_\zeta$. These states 
therefore carry standard charges with respect to the Standard Model 
gauge group. However, they carry non standard charges with respect
to $U(1)_\zeta$. That is, while these states are standard with respect to
the Standard Model, they are exotic with respect to $E_6$. 

The general characteristic of string vacua, due to the breaking
of the non--Abelian GUT symmetries by Wilson lines with a left over
unembedded $U(1)$ symmetry, is the existence of 
massless states that do not satisfy the $U(1)$ quantisation 
of the underlying GUT symmetry \cite{ww}.
In many models the resulting exotic states carry fractional 
electric charge, which are severely constrained by experiments
\cite{vhalyo}. 

A theorem by Schellekens states that any string vacuum 
in which the non--Abelian GUT symmetry is broken by discrete Wilson
lines, necessarily contains states with fractional electric 
charge, 
provided that the weak hypercharge possess the canonical
GUT normalisation \cite{bert}. 
There exist, however, quasi--realistic string models in which the 
exotic fractionally charged states only appear in the
massive spectrum and do not arise at the massless level 
\cite{psclass}. Such exophobic three generation 
models were found when the $SO(10)$ GUT symmetry is 
broken to the Pati--Salam subgroup \cite{psclass}, whereas
models in which the GUT symmetry is broken to the
flipped $SU(5)$ gauge group with odd number of generations 
did not yield any exophobic models \cite{fsu5class}. 
The model arising from the GGSO phases in eq. (\ref{BigMatrix}) 
is an exophobic Pati--Salam model. There are no fractionally 
charged states in the massless spectrum of this string vacuum. 

The $\phi_{1,2}$ and ${\bar\phi}_{1,2}$ states in table 
\ref{tableb} are similarly exotic states. Namely, 
they arise due to the breaking of $E_6$ by discrete 
Wilson lines in the string vacuum. However, as they carry
standard charges with respect to the $SO(10)$ subgroup 
they are not exotic with respect to the Standard Model. 
Such states are therefore a particular signature of the 
string vacuum and may have interesting observational 
consequences \cite{fc}. Furthermore, if they remain
sufficiently light they may be instrumental 
for generating an extended seesaw mechanism \cite{seesaw}.  
It should be remarked, though, that
the $E_6$ exotic states are vector--like 
and are not chiral. Therefore, a priori there is no clear 
argument why they should remain light.

The $\phi_{1,2}$ and ${\bar\phi}_{1,2}$ states fall
into the general category of Wilsonian matter states 
considered in ref. \cite{wilsonian}. Namely, they arise
as a general consequence of the breaking of the $E_6$ 
gauge symmetry by discrete Wilson lines \cite{ww}. 
However, they carry exotic non--$E_6$ charges only with
respect to $U(1)_\zeta$, and consequently with respect to 
$U(1)_{Z^\prime}$, whereas they carry the standard charges 
with respect to the $SO(10)$ subgroup of $E_6$. In fact, 
they are $SO(10)$ singlets. This is quite an 
intriguing situation as it renders them ideal 
dark matter candidates. The reason being that 
if only states with standard $E_6$ charges are used
to break the $U(1)_{Z^\prime}$ symmetry, say the $\chi^\pm$
states and their conjugates, then a local discrete symmetry
\cite{lds} is left which forbids the decay of these states 
to the Standard Model states. The relic abundance 
of such states was considered in ref. \cite{wilsonian}
and it was shown that they may provide viable 
dark matter candidates. However, the singlet
states considered in ref. \cite{wilsonian} 
are Standard Model singlets, but not $SO(10)$ 
singlets. That is they carry exotic charges 
with respect to the $U(1)$ combination, which is 
a combination of $U(1)_{B-L}$ and $U(1)_{T_{3_R}}$, 
and must be broken at a high scale to suppress 
the left--handed neutrino masses. Hence, their
relic abundance depends on an interplay between 
the reheating temperature following 
inflation and the extra $U(1)$ breaking scale. 
However, the states $\phi_{1,2}$ and ${\bar\phi}_{1,2}$
are $SO(10)$ singlets and are not constrained 
by the suppression of the left--handed neutrino 
masses. They may therefore remain light and 
stable down to the $Z^\prime$ breaking scale, 
and may be within reach of forthcoming colliders. 

Table \ref{tablec} contains the vector--like matter 
states that transform non--trivially
under hidden $E_8$ subgroup. All of these states
are $SO(10)$ singlets, but carry nontrivial 
charges under $U(1)_{1,2,3}$. Furthermore, some
of the hidden matter states in \ref{tablec} carry 
exotic charges with respect to $U(1)_\zeta$.
The hidden sector gauge group is broken to $SU(2)^4\times SO(8)$. 
This model may therefore accommodate the self--interacting
dark matter candidates proposed in ref \cite{dm}.

\subsection{The superpotential}

Renormalisable and nonrenormalisable terms in the superpotential
can be calculated by using the tools developed in \cite{kln}.
The cubic level terms in the superpotential are
shown in eq. (\ref{so10sup}--\ref{hidsup}). 
Eq. (\ref{so10sup}) displays the terms that contain
fields that transform nontrivially under the observable 
Pati--Salam group,
\begin{eqnarray}
&
 {\bar{F}_{1R}}\,{F_{1L}}\,{h_1}
+{\bar{F}_{1R}}\,{F_{3L}}\,{h_3}
+{\bar{F}_{1R}}\,{\bar{F}_{2R}}\,{D_4}
+{\bar{F}_{1R}}\,{\bar{F}_{4R}}\,{D_6}
+{F_{1R}}\,{\bar{F}_{3R}}\,{\zeta_1}
\nonumber\\
&
+{\bar{F}_{1R}}\,{\bar{F}_{1R}}\,{D_1}
+{F_{1L}}\,{F_{1L}}\,{D_2}
+{\bar{F}_{3R}}\,{\bar{F}_{3R}}\,{D_2}
+{\bar{F}_{2R}}\,{\bar{F}_{2R}}\,{D_2}
+{F_{2L}}\,{F_{2L}}\,{D_2}
\nonumber\\
&
+{F_{1R}}\,{F_{1R}}\,{\bar{D}_1}
+{F_{3L}}\,{F_{3L}}\,{D_3}
+{\bar{F}_{4R}}\,{\bar{F}_{4R}}\,{D_3}
%\nonumber\\
%&
+{\bar{F}_{3R}}\,{\bar{F}_{4R}}\,{D_7}
+{F_{2L}}\,{F_{3L}}\,{D_7}
\nonumber\\
&
+{h_2}\,{h_2}\,{\Phi_{13}}
+{h_3}\,{h_3}\,{\Phi_{13}}
+{h_1}\,{h_1}\,{\Phi_{12}}
+{h_1}\,{h_2}\,{\chi_5^+}
\nonumber\\
&
+{D_1}\,{D_2}\,{\Phi_{12}}
+{D_2}\,{\bar{D}_1}\,{\bar{\Phi}_{12}^-}
+{D_1}\,{\bar{D}_2}\,{\bar\Phi_{12}^-}
+{\bar{D}_1}\,{\bar{D}_2}\,{\bar\Phi_{12}}
\nonumber\\
&
+{D_2}\,{\bar{D}_3}\,{\Phi_{23}^-}
+{D_2}\,{D_3}\,{\Phi_{23}}
+{D_3}\,{\bar{D}_2}\,{\bar{\Phi}_{23}^-}
+{\bar{D}_2}\,{\bar{D}_3}\,{\bar\Phi_{23}}
\nonumber\\
&
+{D_1}\,{D_3}\,{\Phi_{13}}
+{D_1}\,{\bar{D}_3}\,{\Phi_{13}^-}
+{D_3}\,{\bar{D}_1}\,{\bar{\Phi}_{13}^-}
+{\bar{D}_1}\,{\bar{D}_3}\,{\bar{\Phi}_{13}}
\nonumber\\
&
+{D_4}\,{D_4}\,{\Phi_{12}}
+{D_5}\,{D_5}\,{\Phi_{13}}
+{D_6}\,{D_6}\,{\Phi_{13}}
+{\bar{D}_6}\,{\bar{D}_6}\,{\bar\Phi_{23}}
+{D_7}\,{D_7}\,{\Phi_{23}}
\nonumber\\
&
+{D_5}\,{D_7}\,{\chi_1^+}
+{D_3}\,{D_4}\,{\chi_1^+}
+{D_2}\,{D_5}\,{\chi_2^+}
+{D_2}\,{D_6}\,{\chi_3^+}
+{D_1}\,{\bar{D}_6}\,{\bar{\chi}_4^-}
+{D_1}\,{D_7}\,{\chi_5^+}
\nonumber\\
&
+{\bar{D}_1}\,{\bar{D}_6}\,{\bar{\chi}_4^+}
+{\bar{D}_1}\,{D_7}\,{\chi_5^-}
+{\bar{D}_2}\,{D_5}\,{\chi_2^-}
+{\bar{D}_2}\,{D_6}\,{\chi_3^-}
+{\bar{D}_3}\,{D_4}\,{\chi_1^-}
\nonumber\\
&
+{D_6}\,{\bar{D}_6}\,{\zeta_1}
+{D_4}\,{D_5}\,{\chi_5^+}
+{D_4}\,{D_7}\,{\chi_2^+}.
\label{so10sup}%\nonumber
\end{eqnarray}
A particular 
requirement that we impose on the selected string vacuum 
is the existence of a top quark mass term at the 
cubic level of the superpotential \cite{topc}. 
Such potential term may arise from the couplings to 
$h_1$ and $h_3$ in the first two terms
in eq. (\ref{so10sup}). 

Eq. (\ref{nonabsinsup}) contains only states
that are singlets of the observable and hidden non--Abelian group
factors, 
\begin{eqnarray}
&
{\bar\Phi_{12}}\,{\Phi_{13}^-}\,{\Phi_{23}}
+{\bar\Phi_{12}}\,{\Phi_{23}^-}\,{\Phi_{13}}
+{\bar{\Phi}_{12}^-}\,{\bar{\Phi}_{23}^-}\,{\Phi_{13}^-}
+{\bar{\Phi}_{12}^-}\,{\bar\Phi_{23}}\,{\Phi_{13}}
\nonumber\\
&
+{\bar\Phi_{13}}\,{\bar{\Phi}_{23}^-}\,{\Phi_{12}}
+{\bar\Phi_{13}}\,{\bar\Phi_{12}^-}\,{\Phi_{23}}
+{\bar{\Phi}_{13}^-}\,{\bar\Phi_{12}^-}\,{\Phi_{23}^-}
+{\bar\Phi_{23}}\,{\bar{\Phi}_{13}^-}\,{\Phi_{12}}
\nonumber\\
&
+{\bar\Phi_{12}}\,{\chi_1^-}\,{\chi_1^+}
+{\bar\Phi_{13}}\,{\chi_2^-}\,{\chi_2^+}
+{\bar\Phi_{13}}\,{\chi_3^-}\,{\chi_3^+}
+{\Phi_{23}}\,{\bar{\chi}_4^+}\,{\bar{\chi}_4^-}
+{\bar\Phi_{23}}\,{\chi_5^-}\,{\chi_5^+}
\nonumber\\
&
+{\bar{\Phi}_{12}^-}\,{\zeta_1}\,{\zeta_1}
+{\bar{\Phi}_{12}^-}\,{\zeta_2}\,{\zeta_2}
+{\bar{\Phi}_{12}^-}\,{\zeta_3}\,{\zeta_3}
+{\bar{\Phi}_{13}^-}\,{\zeta_4}\,{\zeta_4}
+{\bar{\Phi}_{13}^-}\,{\zeta_5}\,{\zeta_5}
+{\bar{\Phi}_{13}^-}\,{\zeta_6}\,{\zeta_6}
\nonumber\\
&
+{\bar{\Phi}_{13}^-}\,{\zeta_7}\,{\zeta_7}
+{\bar{\Phi}_{23}^-}\,{\zeta_8}\,{\zeta_8}
+{\bar{\Phi}_{23}^-}\,{\zeta_9}\,{\zeta_9}
+{\bar{\Phi}_{23}^-}\,{\zeta_{10}}\,{\zeta_{10}}
+{\bar{\Phi}_{23}^-}\,{\zeta_{11}}\,{\zeta_{11}}
\nonumber\\
&
+{\bar\Phi_{12}^-}\,{\bar{\zeta}_{1}}\,{\bar{\zeta}_{1}}
+{\bar\Phi_{12}^-}\,{\bar{\zeta}_{2}}\,{\bar{\zeta}_{2}}
+{\bar\Phi_{12}^-}\,{\bar{\zeta}_{3}}\,{\bar{\zeta}_{3}}
+{\Phi_{13}^-}\,{\bar{\zeta}_{4}}\,{\bar{\zeta}_{4}}
+{\Phi_{13}^-}\,{\bar{\zeta}_{5}}\,{\bar{\zeta}_{5}}
+{\Phi_{13}^-}\,{\bar{\zeta}_{6}}\,{\bar{\zeta}_{6}}
\nonumber\\
&
+{\Phi_{13}^-}\,{\bar{\zeta}_{7}}\,{\bar{\zeta}_{7}}
+{\Phi_{23}^-}\,{\bar{\zeta}_{8}}\,{\bar{\zeta}_{8}}
+{\Phi_{23}^-}\,{\bar{\zeta}_{9}}\,{\bar{\zeta}_{9}}
+{\Phi_{23}^-}\,{\bar{\zeta}_{10}}\,{\bar{\zeta}_{10}}
+{\Phi_{23}^-}\,{\bar{\zeta}_{11}}\,{\bar{\zeta}_{11}}
\nonumber\\
&
+{\zeta_1}\,{\chi_3^+}\,{\bar{\chi}_4^+}
+{\bar{\zeta}_{1}}\,{\bar{\zeta}_{7}}\,{\bar{\chi}_4^-}
+{\bar{\zeta}_{1}}\,{\chi_3^-}\,{\zeta_9}
+{\zeta_2}\,{\zeta_4}\,{\chi_5^-}
+{\zeta_3}\,{\zeta_5}\,{\chi_5^-}
\nonumber\\
&
+{\bar{\zeta}_{2}}\,{\chi_2^-}\,{\zeta_{10}}
+{\bar{\zeta}_{3}}\,{\chi_2^-}\,{\zeta_{11}}
+{\chi_1^-}\,{\bar{\zeta}_{4}}\,{\bar{\zeta}_{10}}
+{\chi_1^-}\,{\bar{\zeta}_{5}}\,{\bar{\zeta}_{11}}
\nonumber\\
&
+{\Phi_6}\,{\zeta_1}\,{\bar{\zeta}_1}
%\nonumber\\
%&
+{\zeta_1}\,{\bar{\zeta}_{6}}\,{\zeta_8}
+{\zeta_1}\,{\bar{\zeta}_{7}}\,{\zeta_9}
%\nonumber\\
%&
+{\chi_1^-}\,{\chi_2^-}\,{\chi_5^-}
\nonumber\\
&
+{\bar\Phi_{23}}\,{\phi_1}\,{\phi_1}
+{\bar\Phi_{23}}\,{\phi_2}\,{\phi_2}
+{\Phi_{23}}\,{\bar{\phi}_1}\,{\bar{\phi}_1}
+{\Phi_{23}}\,{\bar{\phi}_2}\,{\bar{\phi}_2}
+{\Phi_2}\,{\phi_1}\,{\bar{\phi}_1}
+{\Phi_2}\,{\bar{\phi}_2}\,{\phi_2}.
\label{nonabsinsup}%\nonumber
\end{eqnarray}
Eq. (\ref{hidsup}) contains fields that transform
nontrivially under the hidden sector group factors,
\begin{eqnarray}
&
{\bar\Phi_{13}}\,{H_{34}^2}\, {H_{34}^2} %{S563- 1}\,{S563- 1}
+{\bar\Phi_{13}}\,{H_{12}^3}\, {H_{12}^3} %{S687- 1}\,{S687- 1}
%\nonumber\\
%&
+{\Phi_{13}}\,{H_{12}^2}\,{H_{12}^2}         %{S541- 1}\,{S541- 1}
+{\Phi_{13}}\,{H_{34}^3}\,{H_{34}^3}         %{S605- 1}\,{S605- 1}
\nonumber\\
&
+{\bar{\Phi}_{23}^-}\,{H_{13}^3}\,{H_{13}^3} %{S676- 1}\,{S676- 1}
+{\bar{\Phi}_{23}^-}\,{H_{34}^5}\,{H_{34}^5} %{S669- 1}\,{S669- 1}
+{\bar{\Phi}_{23}^-}\,{H_{14}^3}\,{H_{14}^3} %{S722- 1}\,{S722- 1}
\nonumber\\
&
+{\Phi_{23}^-}\,{H_{13}^2}\,{H_{13}^2}       %{S620- 1}\,{S620- 1}
+{\Phi_{23}^-}\,{H_{34}^4}\,{H_{34}^4}       %{S633- 1}\,{S633- 1}
+{\Phi_{23}^-}\,{H_{14}^2}\,{H_{14}^2}       %{S643- 1}\,{S643- 1}
\nonumber\\
&
+{\bar\Phi_{12}^-}\,{H_{12}^1}\,{H_{12}^1}  %{S375- 1}\,{S375- 1}
+{\bar\Phi_{12}^-}\,{H_{14}^1}\,{H_{14}^1}  %{S429- 1}\,{S429- 1}
+{\bar\Phi_{12}^-}\,{H_{13}^1}\,{H_{13}^1}  %{S443- 1}\,{S443- 1}
\nonumber\\
&
+{\bar\Phi_{12}^-}\,{H_{34}^1}\,{H_{34}^1}  %{S472- 1}\,{S472- 1}
+{\bar\Phi_{12}^-}\,{H_{23}^1}\,{H_{23}^1}  %{S493- 1}\,{S493- 1}
+{\bar\Phi_{12}^-}\,{H_{24}^1}\,{H_{24}^1}  %{S507- 1}\,{S507- 1}
\nonumber\\
&
+{\chi_2^-}\,{H_{34}^1}\,{H_{34}^5}          %{S472- 1}\,{S669- 1}
+{\chi_3^-}\,{H_{14}^1}\,{H_{14}^3}          %{S429- 1}\,{S722- 1}
+{\chi_3^-}\,{H_{13}^1}\,{H_{13}^3}          %{S443- 1}\,{S676- 1}
%\nonumber\\
%&
+{\bar{\chi}_4^-}\,{Z_5}\,{Z_1}             %{S392- 1}\,{S720- 1}
+{\bar{\phi}_2}\,{H_{12}^1}\,{H_{12}^3}      %{S375- 1}\,{S687- 1}
+{\bar{\phi}_2}\,{H_{34}^1}\,{H_{34}^2}      %{S472- 1}\,{S563- 1}
\nonumber\\
&
+{\Phi_{13}^-}\,{Z_1}\,{Z_1}                %{S720- 1}\,{S720- 1}
+{\bar{\Phi}_{23}^-}\,{Z_2}\,{Z_2}           %{S691- 1}\,{S691- 1}
+{\bar{\Phi}_{23}^-}\,{Z_3}\,{Z_3}           %{S650- 1}\,{S650- 1}
+{\Phi_{13}^-}\,{Z_4}\,{Z_4}                %{S536- 1}\,{S536- 1}
+{\bar\Phi_{12}^-}\,{Z_5}\,{Z_5}            %{S392- 1}\,{S392- 1}
\label{hidsup}%\nonumber\\
\end{eqnarray}

As noted above a VEV that cancels the 
anomalous $U(1)$ $D$--term, which is also $F$--flat
to all orders in the superpotential is given by
the VEV of ${\bar\Phi}_{13}^-$.

\begin{table}[!h]
\noindent
{\scriptsize
\openup\jot
\begin{tabular}{|l|l|c|c|c|c||c|}
\hline
sector&field&$SU(4)\times{SU(2)}_L\times{SU(2)}_R$&${U(1)}_1$&${U(1)}_2$&${U(1)}_3$&$U(1)_\zeta$\\
\hline
$S+b_1$&$\bar{F}_{1R}$&$(\ob{4},\bb{1},\bb{2})$&$\hphantom{+}{1/2}$&$\hphantom{+}0$&$\hphantom{+}0$&$\hphantom{+}{1/2}$\\
$S+b_1+e_3+e_5 %(S400)
$&${F}_{1R}$&$(\bb{4},\bb{1},\bb{2})$&$\hphantom{+}{1/2}$&$\hphantom{+}0$&$\hphantom{+}0$&$\hphantom{+}{1/2}$\\
%
%\hline
$S+b_2$            &${F}_{1L}$&$(\bb{4},\bb{2},\bb{1})$&$\hphantom{+}0$&$\hphantom{+}{1/2}$&$\hphantom{+}0$&$\hphantom{+}{1/2}$\\
$S+b_2+e_1+e_2+e_5 %(S498)
$&${F}_{2L}$&$(\bb{4},\bb{2},\bb{1})$&$\hphantom{+}0$&$\hphantom{+}{1/2}$&$\hphantom{+}0$&$\hphantom{+}{1/2}$\\
$S+b_2+e_1 %(S455)
$&$\bar{F}_{2R}$&$(\ob{4},\bb{1},\bb{2})$&$\hphantom{+}0$&$\hphantom{+}{1/2}$&$\hphantom{+}0$&$\hphantom{+}{1/2}$\\
$S+b_2+e_2+e_5 %(S434)
$&$\bar{F}_{3R}$&$(\ob{4},\bb{1},\bb{2})$&$\hphantom{+}0$&$\hphantom{+}{1/2}$&$\hphantom{+}0$&$\hphantom{+}{1/2}$\\
$S+b_3+e_1+e_2 %(S549)
$&${F}_{3L}$&$(\bb{4},\bb{2},\bb{1})$&$\hphantom{+}0$&$\hphantom{+}0$&$\hphantom{+}{1/2}$&$\hphantom{+}{1/2}$\\
$S+b_3+e_2 %(S613)
$&$\bar{F}_{4R}$&$(\ob{4},\bb{1},\bb{2})$&$\hphantom{+}0$&$\hphantom{+}0$&$\hphantom{+}{1/2}$&$\hphantom{+}{1/2}$\\
%\hline
%
\hline
$S+b_3+x %(S349)
$&${h}_{1}$&$(\bb{1},\bb{2},\bb{2})$&$-{1/2}$&$-{1/2}$&$\hphantom{+}0$&$-1$\\
$S+b_2+x+e_5 %(S697)
$&${h}_{2}$&$(\bb{1},\bb{2},\bb{2})$&$-{1/2}$&$\hphantom{+}0$&$-{1/2}$&$-1$\\
%\hline
$S+b_2+x+e_1+e_2 %(S546))
$&${h}_{3}$&$(\bb{1},\bb{2},\bb{2})$&$-{1/2}$&$\hphantom{+}0$&$-{1/2}$&$-1$\\
%\hline
%\hline
%
\hline
$S+b_3+x+e_1 %(S458)
$&$D_4$&$(\bb{6},\bb{1},\bb{1})$&$-{1/2}$&$-{1/2}$&$\hphantom{+}0$&$-1$\\
&$\chi^+_1$&$(\bb{1},\bb{1},\bb{1})$&$\hphantom{+}{1/2}$&$\hphantom{+}{1/2}$&$\hphantom{+}1$&$+2$\\
&${\chi}_1^-$&$(\bb{1},\bb{1},\bb{1})$&$\hphantom{+}{1/2}$&$\hphantom{+}{1/2}$&$-1$&$\hphantom{+}0$\\
&$\zeta_a, a=2,3$&$(\bb{1},\bb{1},\bb{1})$&$\hphantom{+}{1/2}$&$-{1/2}$&$\hphantom{+}0$&$\hphantom{+}0$\\
&$\bar{\zeta}_a, a=2,3$&$(\bb{1},\bb{1},\bb{1})$&$-{1/2}$&$\hphantom{+}{1/2}$&$\hphantom{+}0$&$\hphantom{+}0$\\
\hline
$S+b_2+x+e_1+e_5  %(S568)
$&$D_5$&$(\bb{6},\bb{1},\bb{1})$&$-{1/2}$&$\hphantom{+}0$&$-{1/2}$&$-1$\\
&$\chi^+_2$&$(\bb{1},\bb{1},\bb{1})$&$\hphantom{+}{1/2}$&$\hphantom{+}1$&$\hphantom{+}{1/2}$&{+}2\\
&${\chi}^-_2$&$(\bb{1},\bb{1},\bb{1})$&$\hphantom{+}{1/2}$&$-1$&$\hphantom{+}{1/2}$&\hphantom{+}0\\
&$\zeta_a, a=4,5$&$(\bb{1},\bb{1},\bb{1})$&$\hphantom{+}{1/2}$&$\hphantom{+}0$&$-{1/2}$&\hphantom{+}0\\
&$\bar{\zeta}_a, a=4,5$&$(\bb{1},\bb{1},\bb{1})$&$-{1/2}$&$\hphantom{+}0$&$\hphantom{+}{1/2}$&\hphantom{+}0\\
\hline
$S+b_2+x+e_2 %(610)
$&$D_6$&$(\bb{6},\bb{1},\bb{1})$&$-{1/2}$&$\hphantom{+}0$&$-{1/2}$&$-1$\\
&$\chi^+_3$&$(\bb{1},\bb{1},\bb{1})$&$\hphantom{+}{1/2}$&$\hphantom{+}1$&$\hphantom{+}{1/2}$&$+2$\\
&${\chi}^-_3$&$(\bb{1},\bb{1},\bb{1})$&$\hphantom{+}{1/2}$&$-1$&$\hphantom{+}{1/2}$&$\hphantom{+}0$\\
&$\zeta_a, a=6,7$&$(\bb{1},\bb{1},\bb{1})$&$\hphantom{+}{1/2}$&$\hphantom{+}0$&$-{1/2}$&$\hphantom{+}0$\\
&$\bar{\zeta}_a, a=6,7$&$(\bb{1},\bb{1},\bb{1})$&$-{1/2}$&$\hphantom{+}0$&$\hphantom{+}{1/2}$&$\hphantom{+}0$\\
\hline
$S+b_1+x+e_3 %(S656)
$&$\bar{D}_6$&$(\bb{6},\bb{1},\bb{1})$&$\hphantom{+}0$&$\hphantom{+}{1/2}$&$\hphantom{+}{1/2}$&$+1$\\
&$\bar{{\chi}}^+_4$&$(\bb{1},\bb{1},\bb{1})$&$-1$&$-{1/2}$&$-{1/2}$&$-2$\\
&$\bar{\chi}^-_4$&$(\bb{1},\bb{1},\bb{1})$&$\hphantom{+}1$&$-{1/2}$&$-{1/2}$&$\hphantom{+}0$\\
&$\zeta_a, a=8,9$&$(\bb{1},\bb{1},\bb{1})$&$\hphantom{+}0$&$\hphantom{+}{1/2}$&$-{1/2}$&$\hphantom{+}0$\\
&$\bar{\zeta}_a, a=8,9$&$(\bb{1},\bb{1},\bb{1})$&$\hphantom{+}0$&$-{1/2}$&$\hphantom{+}{1/2}$&$\hphantom{+}0$\\
\hline
$S+b_1+x+e_5 %(S702)
$&$D_7$&$(\bb{6},\bb{1},\bb{1})$&$\hphantom{+}0$&$-{1/2}$&$-{1/2}$&$-1$\\
&$\chi^+_5$&$(\bb{1},\bb{1},\bb{1})$&$\hphantom{+}1$&$\hphantom{+}{1/2}$&$\hphantom{+}{1/2}$&$+2$\\
&${\chi}^-_5$&$(\bb{1},\bb{1},\bb{1})$&$-1$&$\hphantom{+}{1/2}$&$\hphantom{+}{1/2}$&$\hphantom{+}0$\\
&$\zeta_a, a=10,11$&$(\bb{1},\bb{1},\bb{1})$&$\hphantom{+}0$&$\hphantom{+}{1/2}$&$-{1/2}$&$\hphantom{+}0$\\
&$\bar{\zeta}_a, a=10,11$&$(\bb{1},\bb{1},\bb{1})$&$\hphantom{+}0$&$-{1/2}$&$\hphantom{+}{1/2}$&$\hphantom{+}0$\\
\hline
$S+b_3+x+e_2+e_3$&${\zeta}_{1}$&$(\bb{1},\bb{1},\bb{1})$&$\hphantom{+}{1/2}$&$-{1/2}$&$\hphantom{+}0$&$\hphantom{+}0$\\
%\hline
%
$ %(S445)
$&${\bar\zeta}_{1}$&$(\bb{1},\bb{1},\bb{1})$&$-{1/2}$&$\hphantom{+}{1/2}$&$\hphantom{+}0$&$\hphantom{+}0$\\
%\hline
%
\hline
$S+b_1+x+e_3+e_4+e_6$&${\phi}_{1}$&$(\bb{1},\bb{1},\bb{1})$&$\hphantom{+}0$&$\hphantom{+}{1/2}$&$\hphantom{+}{1/2}$&${+}1$\\
%\hline
%
$
%(S628)
$&${\bar\phi}_{1}$&$(\bb{1},\bb{1},\bb{1})$&$\hphantom{+}0$&${-}{1/2}$&$-{1/2}$&$-1$\\
%\hline
%
\hline
$S+b_1+x+e_4+e_5+e_6$&${\phi}_{2}$&$(\bb{1},\bb{1},\bb{1})$&$\hphantom{+}0$&$\hphantom{+}{1/2}$&$\hphantom{+}{1/2}$&$+1$\\
%\hline
%
$
%(S664)
$&${\bar\phi}_{2}$&$(\bb{1},\bb{1},\bb{1})$&$\hphantom{+}0$&${-}{1/2}$&$-{1/2}$&$-1$\\
%\hline
%
\hline
\end{tabular}
}
\caption{\label{tableb}\it
Twisted matter spectrum (observable sector)  and
$SU(4)\times{SU(2)}_L\times{SU(2)}_R\times{U(1)}^3$ quantum numbers. }
\end{table}

\begin{table}
\noindent
{\scriptsize
\begin{tabular}{|l|l|c|c|c|c||c|}
\hline
sector&field&${SU(2)}^4\times{SO(8)}$&${U(1)}_1$&${U(1)}_2$&${U(1)}_3$&$U(1)_\zeta$\\
\hline
$S+b_3+x+e_4 %(S375)
$&$H_{12}^1$&$(\bb{2},\bb{2},\bb{1},\bb{1},\bb{1})$&$\hphantom{+}{1/2}$&$-{1/2}$&$\hphantom{+}0$&$\hphantom{+}0$\\\hline
$S+b_3+x+e_1+e_4 %(S472)
$&$H_{34}^1$&$(\bb{1},\bb{1},\bb{2},\bb{2},\bb{1})$&$-{1/2}$&$+{1/2}$&$\hphantom{+}0$&$\hphantom{+}0$\\\hline
$S+b_2+x+e_1+e_2+e_6 %(S541)
$&$H_{12}^2$&$(\bb{2},\bb{2},\bb{1},\bb{1},\bb{1})$&$-{1/2}$&$\hphantom{+}0$&$-{1/2}$&$-1$\\\hline
$S+b_2+x+e_1+e_5+e_6 %(S563)
$&$H_{34}^2$&$(\bb{1},\bb{1},\bb{2},\bb{2},\bb{1})$&$+{1/2}$&$\hphantom{+}0$&$+{1/2}$&$+1$\\\hline
$S+b_2+x+e_2+e_6 %(S605)
$&$H_{34}^3$&$(\bb{1},\bb{1},\bb{2},\bb{2},\bb{1})$&$-{1/2}$&$\hphantom{+}0$&$-{1/2}$&$-1$\\\hline
$S+b_1+x+e_3+e_4+e_6 %(S633)
$&$H_{34}^4$&$(\bb{1},\bb{1},\bb{2},\bb{2},\bb{1})$&$\hphantom{+}0$&$-{1/2}$&$+{1/2}$&$\hphantom{+}0$\\\hline
$S+b_1+x+e_4+e_5 %(S669)
$&$H_{34}^5$&$(\bb{1},\bb{1},\bb{2},\bb{2},\bb{1})$&$\hphantom{+}0$&$+{1/2}$&$-{1/2}$&$\hphantom{+}0$\\\hline
$S+b_2+x+e_5+e_6 %(S687)
$&$H_{12}^3$&$(\bb{2},\bb{2},\bb{1},\bb{1},\bb{1})$&$+{1/2}$&$\hphantom{+}0$&$+{1/2}$&$+1$\\\hline
$S+b_3+x+z_1+e_2 %(S429)
$&$H_{14}^1$&$(\bb{2},\bb{1},\bb{1},\bb{2},\bb{1})$&$-{1/2}$&$+{1/2}$&$\hphantom{+}0$&$\hphantom{+}0$\\\hline
$S+b_3+x+z_1+e_2+e_4 %(S443)
$&$H_{13}^1$&$(\bb{2},\bb{1},\bb{2},\bb{1},\bb{1})$&$-{1/2}$&$+{1/2}$&$\hphantom{+}0$&$\hphantom{+}0$\\\hline
$S+b_3+x+z_1+e_1+e_2 %(S493)
$&$H_{23}^1$&$(\bb{1},\bb{2},\bb{2},\bb{1},\bb{1})$&$-{1/2}$&$+{1/2}$&$\hphantom{+}0$&$\hphantom{+}0$\\\hline
$S+b_3+x+z_1+e_1+e_2+e_4 %(S507)
$&$H_{24}^1$&$(\bb{1},\bb{2},\bb{1},\bb{2},\bb{1})$&$-{1/2}$&$+{1/2}$&$\hphantom{+}0$&$\hphantom{+}0$\\\hline
$S+b_1+x+z_1+e_3+e_4+e_5 %(S620)
$&$H_{13}^2$&$(\bb{2},\bb{1},\bb{2},\bb{1},\bb{1})$&$\hphantom{+}0$&$-{1/2}$&$+{1/2}$&$\hphantom{+}0$\\\hline
$S+b_1+x+z_1+e_3+e_5 %(S643)
$&$H_{14}^2$&$(\bb{2},\bb{1},\bb{1},\bb{2},\bb{1})$&$\hphantom{+}0$&$-{1/2}$&$+{1/2}$&$\hphantom{+}0$\\\hline
$S+b_1+x+z_1+e_4+e_5 %(S676)
$&$H_{13}^3$&$(\bb{2},\bb{1},\bb{2},\bb{1},\bb{1})$&$\hphantom{+}0$&$+{1/2}$&$-{1/2}$&$\hphantom{+}0$\\\hline
$S+b_1+x+z_1 %(S722)
$&$H_{14}^3$&$(\bb{2},\bb{1},\bb{1},\bb{2},\bb{1})$&$\hphantom{+}0$&$+{1/2}$&$-{1/2}$&$\hphantom{+}0$\\\hline
%
%From here
$S+b_2+x %(S720)
$&$Z_1$&$(\bb{1},\bb{1},\bb{1},\bb{1},\bb{8}_v)$&$-{1/2}$&$\hphantom{+}0$&$+{1/2}$&$\hphantom{+}0$\\\hline
$S+b_1+x+z_2+e_5+e_6 %(S691)
$&$Z_2$&$(\bb{1},\bb{1},\bb{1},\bb{1},\bb{8}_s)$&$\hphantom{+}0$&$+{1/2}$&$-{1/2}$&$\hphantom{+}0$\\\hline
$S+b_1+x+z_2+e_3+e_6 %(S650)
$&$Z_3$&$(\bb{1},\bb{1},\bb{1},\bb{1},\bb{8}_s)$&$\hphantom{+}0$&$+{1/2}$&$-{1/2}$&$\hphantom{+}0$\\\hline
$S+b_2+x+e_1+e_2+e_5 %(S536)
$&$Z_4$&$(\bb{1},\bb{1},\bb{1},\bb{1},\bb{8}_v)$&$-{1/2}$&$\hphantom{+}0$&$+{1/2}$&$\hphantom{+}0$\\\hline
$S+b_3+x+e_3 %(S392)
$&$Z_5$&$(\bb{1},\bb{1},\bb{1},\bb{1},\bb{8}_v)$&$-{1/2}$&$+{1/2}$&$\hphantom{+}0$&$\hphantom{+}0$\\
\hline
\end{tabular}
}
\caption{\label{tablec}\it Twisted matter spectrum (hidden sector)  and
${SU(2)}^4\times{SO(8)}\times{U(1)}^3$ quantum numbers.}
\end{table}

\section{Conclusions}\label{conclude}

Extensions of the Standard Model by an Abelian gauge symmetry 
are among the most popular cases investigated in studies of 
physics beyond the Standard Model. Extra $U(1)$ symmetries 
arise naturally in Grand Unified Theories with $SO(10)$ and
$E_6$ gauge symmetry. Furthermore, the internal consistency 
conditions of string constructions mandate the existence of additional
gauge symmetries, and may be viewed as a general prediction
of string theory. Indeed, since the mid--eighties many 
authors explored the physics implication of a string inspired extra 
$Z^\prime$ vector boson in collider experiments and astroparticle
observatories. Many of those studies are inspired 
by the heterotic--string, which also admit the appealing
GUT structure, and gave rise to string inspired $Z^\prime$ models
with $E_6$ embedding. Surprisingly, however, the construction 
of explicit string derived models that allow the extra
$U(1)$ symmetry to remain unbroken down to low scales 
has proven to be a difficult challenge. The reason
being that the extra $U(1)$s that arise in string
models could not satisfy the phenomenological constraints
that must be imposed on a viable $U(1)$ symmetry down to
low scales. Some of those constraints being: 
family universality; 
anomaly freedom; 
gauge coupling unification; 
suppressed left--handed neutrino masses. 
The main obstacle being the construction of an 
extra anomaly free $U(1)$ with $E_6$ embedding
of its chiral charges. 

In this paper we constructed a string model that can satisfy these 
phenomenological constraints. The model that we constructed is 
a self--dual model under the spinor--vector duality map that was
observed in free fermionic $Z_2\times Z_2$ orbifolds. 
As a consequence of the self--duality property
the chiral states in the model form complete $E_6$ 
$\mathbf{27}$ representations. However, the gauge symmetry in the effective 
low energy
field theory contains a subgroup of $SO(10)$ and is not enhanced to $E_6$. 
This is possible because the different components of the 
$\mathbf{27}$ are obtained from different fixed points of the 
$Z_2\times Z_2$ toroidal orbifold. 
Consequently, the family universal $U(1)$ may remain unbroken
down to low scales. 

Our three generation Pati--Salam model contains the Higgs fields required
for generating a realistic mass spectrum. The model admits a
mass term at the cubic level of the superpotential that may
generate mass for the heavy fermion family at leading order. 
The massless spectrum of the model is free of exotic fractionally 
charged states. 

Perhaps most tantalising is the appearance in the string model
of vector--like states that carry standard charges with respect
to Standard Model gauge group, but carry non--standard charges
with respect to $U(1)_\zeta$, which descends from $E_6$. 
Such states
%cannot arise in $E_6$ GUT models and 
arise in the
string models due to the breaking of the non--Abelian gauge symmetries
by discrete Wilson lines. Combined observation of the 
extra $E_6$ $U(1)$ symmetry and of the $E_6$ exotic states will
therefore provide strong evidence in favour of a string
construction.
Furthermore, as we discussed in section \ref{model}, 
they provide viable dark matter candidates. Existence of 
a light $Z^\prime$ in these models may
therefore not only be accompanied by the extra $E_6$ states,
required for anomaly cancellation, but by the extra exotic 
states that serve as a distinct signature of the string vacua 
and provide viable dark matter candidates.  

\section{Acknowledgements}

AEF would like to thank the University of Oxford and the
Mainz Institute for Theoretical Physic for hospitality.
AEF is supported in part by STFC under contract ST/G00062X/1. 
This research has been co-financed by the European Union 
(European Social Fund - ESF)
and Greek
national funds through the Operational 
Program ``Education and Lifelong Learning" 
of the National Strategic Reference 
Framework (NSRF) - Research Funding Program: 
THALIS Investing in the society of knowledge 
through the European Social Fund. 

%=========================================================================
%======================== REFERENCES =====================================
%=========================================================================

%\vfill\eject

\bigskip
\medskip

\bibliographystyle{unsrt}

\end{document}